\documentclass[twocolumn, showpacs, amsmath, amssymb, prd]{revtex4}

\usepackage{epsf}

\def\be{\begin{equation}}
\def\ee{\end{equation}}

\begin{document}

\title{Comparing Criteria for Circular Orbits in General Relativity}

\author{Monica L. Skoge$^{1}$} 

\author{Thomas W. Baumgarte$^{1,2}$}

\affiliation{$^{1}$Department of Physics and Astronomy, Bowdoin College,
        Brunswick, ME 04011}

\affiliation{$^{2}$Department of Physics, University of Illinois at
       Urbana-Champaign, Urbana, IL, 61801}

\begin{abstract}
We study a simple analytic solution to Einstein's field equations
describing a thin spherical shell consisting of collisionless
particles in circular orbit.  We then apply two independent criteria
for the identification of circular orbits, which have recently been
used in the numerical construction of binary black hole solutions, and
find that both yield equivalent results.  Our calculation illustrates
these two criteria in a particularly transparent framework and
provides further evidence that the deviations found in those numerical
binary black hole solutions are not caused by the different criteria
for circular orbits.
\end{abstract}


\maketitle

Binary black holes are among the most promising sources of
gravitational radiation for the new generation of gravitational wave
detectors LIGO, VIRGO, GEO and TAMA.  Motivated by the need of
theoretical models for the identification and interpretation of
future gravitational wave signals, several researchers have solved the
constraint equations of Einstein's field equations to construct
initial data describing binary black holes in quasi-circular orbit
\cite{c94,b00,ptc00,ggb02,b02,tbcd02}.

Constructing such initial data requires making several choices,
including the decomposition of the initial value problem and the
background geometry and topology.  Moreover, solving the constraint
equations provides the gravitational fields for black holes with
arbitrary separation and momenta, and an additional criterion has to
be applied to identify circular orbits.  It is not surprising that
different choices lead to physically different data.  While all of
these different data may be correct solutions to the constraint
equations of general relativity, some may be more relevant
astrophysically than others, in that they better represent a binary
black hole system as it arises from inspiral from large separation.

The results of Cook \cite{c94} and Baumgarte \cite{b00} (which we will
jointly refer to as CB) and Grandcl\'ement, Gourgoulhon and Bonazolla
\cite{ggb02} (hereafter GGB) differ by about a factor of two in the
orbital frequency for the innermost stable circular orbit.  This
discrepancy raises two questions, namely which results are more
relevant astrophysically and which choice in the respective approaches
are responsible for the deviations.  The better agreement of the GGB
results with post-Newtonian results \cite{dgg02} suggests that
these represent binary black holes in circular orbits more accurately
\cite{footnote1}.  There is also increasing evidence that the
differences between CB and GGB are related to the different
decompositions of the constraint equations \cite{pct02,py02}.  CB
adopt the conformal transverse-traceless decomposition, which allows
for an analytic solution of the momentum constraint \cite{by80}, while
GGB adopt the conformal thin-sandwich decomposition
\cite{wm95,y99,c00} (see also \cite{py02}).  It has been demonstrated
that the two decompositions may lead to physically different data
\cite{pct02}, and it has also been suggested that the thin-sandwich
decomposition together with maximal slicing may provide a more natural
framework for constructing quasi-equilibrium solutions \cite{c00}.

In this Brief Report we explore the effect of another difference in
the approaches of CB and GGB, namely the criterion for locating
circular orbits.  CB adopt a turning-point method, in which circular
orbits are identified with extrema of the binding energy (see
eq.~(\ref{tp_crit}) below), while GGB identify circular orbits by
equating the ADM \cite{adm62} and Komar masses \cite{k59}
(eq.~(\ref{m_crit})).  Since the two mass definitions agree only for
stationary spacetimes, this criterion is closely related to imposing a
relativistic virial theorem \cite{bg94}.

To explore the effect of these different criteria for circular orbits
in a particularly simple and transparent framework, we apply them to
an analytic solution of Einstein's equations describing a thin,
spherical shell of identical collisionless particles.  At every point
on the shell the particles move isotropically, but all with the same
speed in the plane perpendicular to the radius.  In an oscillating
shell, each particle moves about the center in a bound orbit.  In the
Newtonian limit, each orbit is a closed ellipse, and for static shells
each orbit is circular (compare \cite{ybs01}).  Since each particle
follows a geodesic, circular orbits can be identified without
ambiguity.  These orbits can then be compared with those obtained from
the turning-point and mass methods.

In the following we will focus on a moment of time symmetry, when at
least momentarily each particle is in a purely tangential orbit $u^r =
0$ (where $u^a$ is the four-velocity).  The spherically symmetric line
element can then be written as
\begin{equation} \label{metric}
ds^2 = -\alpha^2 dt^2 + \psi^4 (dr^2 + r^2 (d \theta + \sin^2 \theta
d\phi)),
\end{equation}
where $\alpha$ is the lapse function and $\psi$ the conformal factor.

The rest mass $M_0$ of the shell can be computed from
\begin{equation}
M_0 = \int \rho_0 u^t \sqrt{-g} d^3x
= 4 \pi \int \rho_0 W \psi^6 r^2 dr,
\end{equation}
where $g$ is the determinant of the spacetime metric and where we have
defined the particles' Lorentz factor $W \equiv - \alpha u^t$.  Since
the shell's co-moving density $\rho_0$, which is a sum of the
individual particle densities $\rho_0^A$, vanishes everywhere except
at the radius $R$ of the shell, we find
\begin{equation}
\rho_0 = \sum_A \rho_0^A = 
\frac{M_0}{4 \pi R^2 W \psi^6} \, \delta(r - R).
\end{equation}

The conformal factor $\psi$ in (\ref{metric}) can now be found from 
the Hamiltonian constraint
\begin{equation} \label{ham1}
\nabla^2 \psi = -2 \pi \psi^5 \rho_N, 
\end{equation} 
and, following GGB, the lapse $\alpha$ from the maximal slicing
condition
\begin{equation} \label{maxslicing} 
\nabla^2 (\alpha \psi) = 2 \pi \alpha \psi^5 (\rho_N + 2S).
\end{equation} 
Here $\rho_N$ is the density measured by a normal observer $n^a$ 
\begin{equation}
\rho_N = n^a n^b T_{ab} 
= n^a n^b \sum_A \rho_0^A u_a^A u_b^A
= \rho_0 W^2,
\end{equation}
(compare \cite{djs00,st}), and $S$ is the trace of the spatial stress
\begin{equation} 
S = \gamma^{ij} T_{ij} = \rho_0 \gamma^{ij} u_i u_j = \rho_0 (W^2 - 1),
\end{equation}
where we have used the normalization condition
\begin{equation} \label{norm}
1 = W^2 - \gamma^{ij} u_i u_j.
\end{equation}

For time symmetry both the momentum density $j^a = - \gamma^{ab} n^c
T_{bc}$ and the extrinsic curvature vanish, so that a zero shift
$\beta^i = 0$ identically satisfies the shift equation obtained in the
conformal thin-sandwich decomposition.

The Hamiltonian constraint (\ref{ham1}) and the maximal slicing condition
(\ref{maxslicing}) can readily be solved analytically by matching 
two vacuum solutions at the shell's radius $R$.  Choosing the vacuum
solutions such that the interior solution is regular at the center,
while the exterior solution is regular at infinity, we find for the
conformal factor
\begin{equation} \label{cf1}
\psi = \left\{ \begin{array}{ll}
	\displaystyle 1 + \frac{W}{2\psi|_{\bar R} \bar R} & 
		\mbox{~~~for~~} 0 \leq \bar r < \bar R  
	\\[3mm]
	\displaystyle 1 + \frac{W}{2\psi|_{\bar R} \bar r} & 
		\mbox{~~~for~~} \bar r \geq \bar R.  
\end{array}
\right.
\end{equation}
Here and in the following we non-dimensionalize all quantities with
respect to $M_0$, e.g.~$\bar r \equiv r/M_0$.  The value of $\psi|_{\bar R}$
can be found by evaluating the conformal factor at $\bar r = \bar R$,
which yields a quadradic equation with the solution
\begin{equation}
\label{psir}
\psi|_{\bar R} = \frac{1}{2} + \sqrt{\frac{1}{4} + \frac{W}{2 \bar R}}.
\end{equation}
The sign has been chosen so that $\psi$ approaches the gravitational
potential $\phi_{\rm Newt}$ in the Newtonian limit.

In terms of the ADM mass \cite{adm62,my74} 
\begin{equation} \label{m_adm}
\bar M_{\rm ADM} = - \frac{1}{2\pi M_0 } \oint_\infty D^i \psi d^2S_i
 	= \frac{W}{\psi|_{\bar R}},
\end{equation}
the exterior conformal factor (\ref{cf1}) can be written
\begin{equation} \label{cf2}
\psi = 	1 + \frac{\bar M_{\rm ADM}}{2 \bar r} 
\mbox{~~~for~~}\bar r \geq \bar R. 
\end{equation}

The maximal slicing condition (\ref{maxslicing}) can be solved
analogously to the Hamiltonian constraint, yielding
\begin{equation}
\alpha \psi = \left\{ \begin{array}{ll} \displaystyle 1 -
	\frac{\alpha|_{\bar R} (3W^2-2)}{2W\psi|_{\bar R}\bar R} &
	\mbox{~~~for~~} 0 \leq \bar r < \bar R \\[3mm] \displaystyle 1
	- \frac{\alpha|_{\bar R} (3W^2-2)}{2W\psi|_{\bar R}\bar r}
	& \mbox{~~~for~~} \bar r \geq \bar R.
\end{array}
\right.
\end{equation}
Dividing by $\psi$, we find in the exterior 
\begin{equation}
\alpha = \frac{-\alpha|_{\bar R} (3W^2-2)+2W\psi|_{\bar R}\bar r}{W^2+2W\psi|_{\bar R}\bar r}
\mbox{~~~for~~}\bar r \geq \bar R.
\end{equation}
Evaluating this expression at $\bar r = \bar R$ determines the coefficient 
$\alpha|_{\bar R}$
\begin{equation}
\alpha|_{\bar R} = \left( 1+\frac{2W^2-1}{\psi|_{\bar R} \bar R W}
\right)^{-1}.
\end{equation}

Following GGB we now compute the Komar mass \cite{k59}
\begin{equation} \label{m_komar}
\bar M_{\rm K} = \displaystyle\frac{1}{4\pi M_0} 
	\oint_\infty D^i \alpha \, d^2 S_i 
= \frac{\alpha|_{\bar R} (3W^2-2) + W^2}{2W \psi|_{\bar R}}.
\end{equation}
We note that the Komar mass is a slicing dependent quantity, and that
this particular form results from having imposed maximal slicing.  In
terms of the Komar and ADM masses, the exterior lapse $\alpha$ can be
written
\begin{equation}
\alpha = \frac{2 \bar r - (2\bar M_{\rm K}-\bar M_{\rm ADM})}
	{2\bar r + \bar M_{\rm ADM}}
\mbox{~~~for~~}\bar r \geq \bar R. 
\end{equation}
This expression reduces to the lapse as identified from the
Schwarzschild metric in isotropic coordinates (see, e.g., exercise
31.7 in \cite{mtw73}) only if the two masses agree, $\bar M_{\rm K} =
\bar M_{\rm ADM}$ (compare criterion (\ref{m_crit}) below).

So far, the shell's radius $\bar R$ and Lorentz factor $W$ appear
independently in the above equations.  It is intuitively clear that
searching for circular orbits will yield a relation between the
particles' angular velocity and the gravitational field, and hence
between $\bar R$ and $W$.  Since our model consists of collisionless
particles, circular orbits can be determined directly by solving
the geodesic equations.  Since all particles are identical, it is
sufficient to evaluate the equation of motion for one particle, which
we take to orbit in the equatorial plane.  We therefore have
$u^{\theta} = u^r = 0$, so that the normalization condition
(\ref{norm}) yields a relation between $u^{\phi}$ and $W$
\begin{equation} \label{uphi}
(u^{\phi})^2 = \frac{W^2-1}{\psi^4|_R R^2},
\end{equation}
where we temporarily drop the bar notation.  We now evaluate the
geodesic equation,
\begin{equation}
\frac{d u^a}{d\lambda} + \Gamma^a_{bc} u^b u^c = 0,
\end{equation}
for $a = r$ to find a condition for the particles to remain in 
a purely tangential orbit ($d u^r/d \lambda = 0$)
\begin{equation} \label{gam}
\Gamma^r_{tt}(u^t)^2 + \Gamma^r_{\phi \phi}(u^\phi)^2 = 0.
\end{equation}
Combining this with (\ref{uphi}) and $W = \alpha u^t$ gives
\begin{equation} \label{w2}
W^2 = \left(1+ \frac{\psi^4|_R R^2 \Gamma^r_{tt}}
	{\alpha^2 \Gamma^r_{\phi \phi}}\right)^{-1}.
\end{equation}
When evaluating the Christoffel symbols, we must take into account the
discontinuity in the first derivative of the metric coefficients at $r
= R$.  By averaging such a quantity over an extended shell and letting
the thickness of the shell go to zero, we find that the derivative has
to be replaced with
\begin{equation}
\psi_{,r} \rightarrow \frac{1}{2}(\psi_{,r}|_+ + \psi_{,r}|_-) =
\frac{1}{2} \psi_{,r}|_+.
\end{equation}
Using this rule for both $\psi$ and $\alpha$ we find
\begin{equation}
\Gamma^r_{\phi \phi} = \displaystyle
\frac{M_{\rm ADM}}{2}\left(1+\displaystyle\frac{M_{\rm ADM}}{2R}\right)^{-1} 
	- R 
\end{equation}
and
\begin{equation}
\Gamma^r_{tt} = \displaystyle
\frac{M_{\rm K}}{2R^2}\left(1 + \frac{M_{\rm ADM}}{2R}\right)^{-6} 
\left(1 - \frac{M_{\rm K}}{R + M_{\rm ADM}/2} \right)
\end{equation}
at $r=R$.  Inserting these into eq.~(\ref{w2}) yields 
\begin{equation} \label{geo}
W^2 = \left(1-\frac{M_{\rm K}}{2R -2M_{\rm K} + M_{\rm ADM}}\right)^{-1}.
\end{equation}
After some algebraic manipulation and dividing out the unphysical root
$W=0$, eq. (\ref{geo}) can be expanded into
\begin{equation} \label{5p}
4 W^5 - 6\bar R W^4 - 4 W^3 + 10 \bar R W^2 + W - 4 \bar R = 0,
\end{equation} 
where we have reintroduced the bar notation.  This is the condition
relating $W$ and $\bar R$ for circular orbits.  It is easy to show
that this equation reduces to $\bar \Omega^2 = \bar R^{-3}/2$ in the
Newtonian limit (with $\bar R \gg 1$, $v \ll 1$ and $W \simeq 1 +
v^2/2 = 1 + \bar R^2 \bar \Omega^2/2$).

For black holes, alternative criteria have to be used to identify
circular orbits.  In the following we will compare the turning-point
method adopted by CB and the mass criterion adopted by GGB.

In the turning-point method, a circular orbit is identified by finding
an extremum of the ADM mass (or equivalently the binding energy) at
constant angular momentum $\bar u_{\phi}$
\begin{equation} \label{tp_crit}
\left. \frac{d \bar M_{\rm ADM}}{d \bar R} \right|_{\bar u_{\phi}} = 0.
\end{equation}
In a Newtonian context, this condition arises naturally from
Hamilton's equations of motion.  We start by differentiating the
normalization condition, $(\bar u_\phi)^2 = \psi^4|_{\bar R} M_0^2 \bar R^2
(W^2-1)$, with respect to $\bar R$ to find
\begin{equation} \label{dw}
\frac{dW}{d \bar R} = \frac{-(W^2-1)(1+b)}
	{\bar RW(1+b)+4W^2-2}
\end{equation}
for sequences of constant angular momentum, where for convenience we
have abbreviated $b = (1 + 2 W/\bar R)^{1/2}$.  We now
locate an extremum of the ADM mass (\ref{m_adm}) by setting its
derivative with respect to $\bar R$ equal to zero
\begin{equation} \label{dw2}
\frac{dW}{d\bar R}\left(\frac{W}{\bar R b(1+b)}-1\right) 
	= \frac{W^2}{\bar R^2 b(1+b)}.
\end{equation}
Combining (\ref{dw}) and (\ref{dw2}) then yields the condition
\begin{equation} \label{tp}
W^2 = \frac{-\bar R(W^2-1)(1+b)(W-\bar R b(1+b))}{\bar RW(1+b)+4W^2-2}.
\end{equation}
Inserting $b$ and eliminating the unphysical root $W = -\bar R/2$, 
eq. (\ref{tp}) can be expanded identically into eq.~(\ref{5p}).  

In the mass method of GGB, the condition for circular orbits is
obtained by equating the ADM and Komar mass (as obtained from maximal
slicing)
\begin{equation} \label{m_crit}
\bar M_{\rm ADM} = \bar M_{\rm K}.
\end{equation}
Inserting (\ref{m_adm}) and (\ref{m_komar}) yields, after some
manipulation and elimination of the unphysical root $W=0$, again the
condition (\ref{5p}).  Thus we have established that both criteria
yield the correct condition for circular orbits in our model problem.

Since (\ref{m_crit}) only holds for stationary spacetimes, this
criterion is closely related to a relativistic virial theorem.  This
relation is also evident from the expansions of the ADM and Komar
masses to first order in $\epsilon \sim 1/\bar R \sim v^2$,
\begin{equation} \label{m_adm_newt}
\bar M_{\rm ADM} \simeq 
1 - \frac{1}{2 \bar R} + \frac{1}{2} v^2 = 1 + \bar U + \bar T
\end{equation}
and
\begin{equation} \label{m_komar_newt}
\bar M_{\rm K} \simeq 1 - \frac{1}{\bar R} + \frac{3}{2} v^2 =
1 + 2 \bar U + 3 \bar T,
\end{equation}
where $U$ and $T$ are the Newtonian potential and kinetic energies of
the spherical shell.  The two expansions (\ref{m_adm_newt}) and
(\ref{m_komar_newt}) are equal only if the Newtonian virial theorem $T
= - U/2$ holds.  

For completeness, we evaluate the relativistic virial theorem in
spherical symmetry as derived by \cite{bg94}
\begin{equation}
\int \left(4\pi S - \frac{1}{\psi^4}\left( \big(\frac{d\ln \alpha}{dr}
\big)^2 -\frac{1}{2} \big(\frac{d\ln \psi^2}{dr}
\big)^2\right)\right)\psi^6 r^2 dr = 0.
\end{equation}
Computing the above integral in terms of the Komar and ADM masses yields
\begin{equation}
\frac{(W^2-1)}{W} - 
\frac{2\bar M_{\rm K}^2}{\bar M_{\rm ADM}-2\bar M_{\rm K}+2\bar R} + 
\frac{\bar M_{\rm ADM}^2}{2 \bar R} = 0,
\end{equation}
which can again be brought into the form (\ref{5p}).

We now briefly discuss the physical implications of the condition
(\ref{5p}).  Solving for $\bar R$ we find
\begin{equation}
\bar R = \frac{4 W^5 - 4 W^3 + W}{6 W^4 - 10 W^2 + 4}.
\end{equation}
To find a minimum value for the radius of our shell, we extremize the
above equation with respect to $W$, which yields
\begin{equation}
(2W^2 - 1)(6W^6 - 21W^4 + 15W^2 - 2) = 0.
\end{equation}
The only physical root (i.e.~$W$ real and $W \geq 1$) is $W = 1.607$,
corresponding to $\bar R_{\rm min} = 1.532$.  Expressing this in
terms of $M_{\rm ADM}$ and circumferential radius $R_{\rm C}$ we
find 
\begin{equation}
\left( \frac{R_{\rm C}}{M_{\rm ADM}} \right)_{\rm min} = 2.506
\mbox{~~~~(equilibrium)}.
\end{equation}
This value should be compared with the Buchdahl limit $(R_{\rm
C}/M_{\rm ADM})_{\rm min} = 9/4 = 2.25$ \cite{b59} for static fluid
balls and $(R_{\rm C}/M_{\rm ADM})_{\rm min} = 3$ \cite{mtw73} for
test particles in circular orbit in Schwarzschild spacetimes.

Requiring the particles' orbits to be stable leads to a more stringent
limit on the compaction, which we find by requiring the second
derivative of $ \bar M_{\rm ADM}$ with respect to $\bar R$ to vanish
in addition to (\ref{tp_crit}).  This yields an equation for $W$ with
the physical root $W=1.108$ corresponding to $\bar R_{\rm min} =
3.053$, or
\begin{equation}
\left( \frac{R_{\rm C}}{M_{\rm ADM}} \right)_{\rm min} = 4.265
\mbox{~~~~(stability)},
\end{equation}
which should be compared with the innermost stable circular orbit
$(R_{\rm C}/M_{\rm ADM})_{\rm min} = 6$ of test particles in
Schwarzschild spacetimes.

To summarize, we construct an analytic solution to Einstein's field
equations describing a thin spherical shell consisting of
collisionless particles in circular orbits.  We apply the
turning-point criterion (\ref{tp_crit}) used by CB and the mass
criterion (\ref{m_crit}) used by GGB and find that both conditions
correctly identify circular orbits.  The later criterion is intimately
related to adopting maximal slicing, which is a natural choice for
constructing quasi-equilibrium spacetimes (compare \cite{c00}).  Our
calculation illustrates these two criteria in the context of a very
transparent, analytical framework and provides further evidence that
the differences between the findings of CB and GGB result from the
different initial value decompositions.

\acknowledgments
 
MLS gratefully acknowledges support through the Surdna Foundation
Undergraduate Research Fellowship Program.  We would also like to
thank R.~H.~Price and K.~S.~Thorne for useful conversations, as well
as the Visitors Program in Numerical Relativity at Caltech, where this
project was initiated, for extending their hospitality.  This work was
supported in part by NSF Grant PHY 01-39907 to Bowdoin College.


\begin{thebibliography}{99}
\bibitem{footnote1} In principle it is possible that the choices and
approximations made in GGB and the post-Newtonian calculations
(e.g.~\cite{djs00,bd99}) lead to similar errors, so that their
agreement might be misleading.  Fully self-consistent numerical
relativity evolution calculations would therefore provide a more
reliable indication as to which data-sets indeed lead to circular
orbits.
\end{thebibliography}
\end{document}